\documentclass[11pt]{article}

\usepackage[utf8]{inputenc}
\usepackage[T1,tone,safe]{tipa}

\usepackage{latexsym,amsmath,amssymb,stmaryrd,theorem,epsfig}

\usepackage{pifont}
\usepackage{array}

\usepackage{setspace}
\usepackage{url}
\usepackage{multirow}
\usepackage[colorlinks=true,linkcolor=blue,urlcolor=blue,citecolor=blue,linktoc=page]{hyperref}

\usepackage{mciteplus}

\mciteSetMidEndSepPunct{\nolinebreak;\space}{\nolinebreak.}{\relax}

\DeclareFontFamily{OT1}{rsfs10}{}
\DeclareFontShape{OT1}{rsfs10}{m}{n}{ <-> rsfs10 }{}
\DeclareMathAlphabet{\mathscript}{OT1}{rsfs10}{m}{n}

\numberwithin{equation}{section}

\newcommand{\tr}{\text{tr}}
\newcommand{\cc}[1]{{#1}^*}

\newcommand{\dif}{\mathrm{d}}

\newcommand{\R}{\mathbb{R}}

\newcommand{\hv}{\mathrm{h}}

\newcommand{\alg}[1]{\mathfrak{#1}}

\def \be {\begin{equation}}
\def \ee {\end{equation}}

\def \tr {{\rm tr}}

\def \hh {{\rm h}}

\def \iffa {\iffalse}

\def \beq {\begin{eqnarray}}
\def \eeq {\end{eqnarray}}

\def \bea {\begin{align}}
\def \eea {\end{align}}

\def \ww {w} 
\def \qhat {\kappa} 

\usepackage[normalem]{ulem}

\begin{document}

\overfullrule=0pt
\parskip=3pt
\parindent=12pt

\vspace{ -3cm}
\thispagestyle{empty}
\vspace{-1cm}

\rightline{ Imperial-TP-AS-2014-01}

\begin{center}
\vspace{1cm}
{\Large\bf String theory in $AdS_3 \times S^3 \times T^4$ with mixed flux:
\\ \vspace{0.1cm}
semiclassical and 1-loop phase in the S-matrix
\vspace{0.1cm} }
\vspace{1.0cm}

{A. Stepanchuk$^{a,}$\footnote{a.stepanchuk11@imperial.ac.uk}}\\

\vskip 0.3cm

{\em $^{a}$ The Blackett Laboratory, Imperial College, London SW7 2AZ, U.K.}

\vspace{.2cm}

\end{center}

\baselineskip 11pt
\begin{abstract}
\noindent
We present a semiclassical derivation of the tree-level and 1-loop dressing phases in the massive sector of string theory on $AdS_3 \times S^3 \times T^4$ supplemented by R-R and NS-NS 3-form fluxes. In analogy with the $AdS_5\times S^5$ case, we use the dressing method to obtain scattering solutions for dyonic giant magnons which allows us to determine the semiclassical bound-state S-matrix and its 1-loop correction. We also find that the 1-loop correction to the dyonic giant magnon energy vanishes. Looking at the relation between the bound-state picture and elementary magnons in terms of the fusion procedure we deduce the elementary dressing phases. In both the semiclassical and 1-loop cases we find agreement with recent proposals from finite-gap equations 
and unitarity cut methods. 
Further, we find consistency with the finite-gap picture by determining the resolvent for the dyonic giant magnon from the semiclassical bosonic scattering data.
\end{abstract}

\newpage

\setcounter{equation}{0}
\setcounter{footnote}{0}
\setcounter{section}{0}

\tableofcontents

\baselineskip 14pt

\newcommand{\Li}{\operatorname{Li}}
\newcommand{\sech}{\operatorname{sech}}
\def \hh {{\rm h}}

\section{Introduction}\label{sec:int}

In the framework of the AdS/CFT correspondence integrability has allowed for significant progress in understanding $AdS_5/CFT_4$, the duality between $\mathcal{N}=4$ Super-Yang-Mills and type IIB string theory on $AdS_5\times S^5$ and $AdS_4/CFT_3$, the duality between ABJM Super Chern-Simons and type IIA string theory on $AdS_4\times CP^3$ \cite{Beisert:2010jr}. These theories involve 32 and 24 supersymmetries respectively and a natural question is whether we can further extend integrability techniques to similar but less (super)symmetric settings. Recently there has been progress in this direction for $AdS_3/CFT_2$ with the maximally supersymmetric backgrounds of $AdS_3\times S^3\times T^4$ and $AdS_3 \times S^3 \times S^3 \times S^1$, which preserve 16 supersymmetries. These backgrounds can be further supplemented with Ramond-Ramond (R-R) and Neveu-Schwarz-Neveu-Schwarz (NS-NS) 3-form fluxes.\footnote{All other fluxes are zero and the dilaton is constant.} Physically $AdS_3$ backgrounds are also of interest as they are related to BTZ black hole solutions \cite{Banados:1992wn}.\\
\\
In this paper we consider string theory on $AdS_3\times S^3\times T^4$. Its type IIB supergravity solutions with pure R-R, pure NS-NS and mixed flux arise as the near-horizon limits of the D5-D1, the NS5-NS1 and the D5-D1 + NS5-NS1 brane systems \cite{Giveon:1998ns}. These R-R and NS-NS solutions are related by S-duality and, in particular, we can obtain the mixed flux background by applying S-duality to either the pure R-R or the pure NS-NS solution. Thus, string theory on the mixed flux background also provides an important model to study the non-perturbative S-duality.\\
\\
In the pure NS-NS theory the free string spectrum can be found using a chiral decomposition of its formulation as a supersymmetric extension of an $SL(2,R)\times SU(2)$ WZW model \cite{Maldacena:2000hw,*Maldacena:2000kv,*Maldacena:2001km}. In the pure R-R case there is no equivalent of this technique. Instead the exact spectrum is believed to be described in terms of an integrability-based approach in analogy to the $AdS_5\times S^5$ case \cite{Babichenko:2009dk,OhlssonSax:2011ms,Borsato:2012ud,Borsato:2012ss,Borsato:2013qpa,Borsato:2013hoa,Borsato:2014exa,Borsato:2014hja}. However, the mixed flux theory connects these seemingly distinct cases. Its R-R and NS-NS 3-form fluxes are given by (choosing unit curvature radii)
\begin{align}
F = \qhat \Big(\mathrm{vol}(AdS_3) + \mathrm{vol}(S^3)\Big),\qquad H = q \Big(\mathrm{vol}(AdS_3) + \mathrm{vol}(S^3)\Big).
\end{align}
with their coefficients related by the supergravity equations as
\begin{align}
q^2 + \qhat^2 = 1.
\end{align}
This provides us with an interpolating theory between the pure NS-NS theory at $q=0$ and the pure R-R theory at $q=1$. Therefore, solving for the spectrum in the mixed flux case should improve our understanding of the connection between the world-sheet CFT methods and the integrability-based approach.\\
\\
At the classical level integrability of a theory can be established by finding a Lax representation for its equations of motion. The Lax connection is then associated with a monodromy matrix which generates an infinite tower of conserved quantities of the integrable field theory. Such a Lax construction has been achieved for string $\sigma$-models on semi-symmetric coset spaces such as $AdS_5\times S^5$ \cite{Metsaev:1998it}, as well as R-R flux supported backgrounds including $AdS_4\times CP^3$ \cite{Arutyunov:2008if,*Stefanski:2008ik} and the above maximally supersymmetric $AdS_3$ backgrounds \cite{Babichenko:2009dk,Zarembo:2010sg} with their supercosets
\begin{align}
%
AdS_3\times S^3 \simeq \frac{PSU(1,1|2)\times PSU(1,1|2)}{SU(1,1)\times SU(2)},\qquad
AdS_3\times S^3\times S^3 \simeq \frac{D(2,1;\alpha)\times D(2,1;\alpha)}{SU(1,1)\times SU(2)\times SU(2)}.
\end{align}
In the mixed flux theory classical integrability and UV finiteness have been shown by supplementing this construction with an additional Wess-Zumino (WZ) term for the NS-NS flux in the action \cite{Babichenko:2009dk,Cagnazzo:2012se}. This integrability is also expected to extend to the quantum level, leading to the possibility of determining the exact string spectrum using a thermodynamic Bethe ansatz (TBA) and improving our understanding of the corresponding CFT dual. For a general review on integrability in $AdS_3/CFT_2$ also see \cite{Sfondrini:2014via}.\\
\\
An essential ingredient in solving for the string spectrum using integrability methods is the two-particle S-matrix for the scattering of fundamental world-sheet excitations. In an integrable theory this S-matrix is constrained, up to overall scalar factors - the dressing phases, by the Yang-Baxter equation (YBE). However, while in $AdS_5/CFT_4$ and $AdS_4/CFT_3$ all world-sheet excitations are massive, a novel feature of $AdS_3/CFT_2$ is the presence of additional massless excitations from the $T^4$ and $S^1$ directions. Incorporating these massless modes into the well established integrability framework has posed some challenges. In particular, their relativistic treatment requires a more abstract notion of an S-matrix \cite{Zamolodchikov:1992zr}. Recently there has been progress in resolving this issue by treating massless and massive modes in a common non-relativistic integrability framework and the complete world-sheet S-matrix has been obtained in the case of $AdS_3\times S^3\times T^4$ with mixed flux \cite{Borsato:2014exa,Borsato:2014hja,Hoare:2013pma,Lloyd:2014bsa}.\\
\\
In this paper we are interested in the dressing phase for the massive sector of the mixed flux theory. Originally in the case of $AdS_5\times S^5$ the semiclassical (AFS) and 1-loop (HL) dressing phases were established using a Bethe ansatz in \cite{Arutyunov:2004vx,Hernandez:2006tk}. They were reproduced from a first-principle semiclassical soliton quantisation approach for giant magnon solutions in \cite{Chen:2006gq,*Roiban:2006gs,Chen:2007vs} and later in \cite{Beisert:2006ez} also the all-loop (BES) dressing phase was found.\\
\\
Also in the case of $AdS_3 \times S^3 \times T^4$ without flux these dressing phase factors have been established at tree-level and 1-loop orders using semiclassical methods in \cite{David:2010yg,Beccaria:2012kb,Abbott:2013ixa} and an all-loop proposal was given in \cite{Borsato:2013hoa} by solving the crossing equations satisfied by the dressing phase \cite{Borsato:2013qpa,Borsato:2014exa,Borsato:2014hja}. In \cite{Beccaria:2012kb} the 1-loop phase was further constrained in the pure R-R case by considering quantum corrections to spinning strings.\\
\\
In the massive sector of $AdS_3 \times S^3 \times T^4$ with mixed flux results for the semiclassical and 1-loop phases have been obtained using the approach of finite-gap equations and algebraic curve quantisation in \cite{Babichenko:2014yaa} and matching conjectures for the 1-loop phases have also been proposed using unitarity cut based methods in \cite{Engelund:2013fja,*Bianchi:2014rfa}. For the semiclassical phase the proposal in \cite{Babichenko:2014yaa} relies on the assumption that the dressing phase for the scattering of same-type excitations remains unchanged when switching on the flux and therefore is given by the usual AFS expression \cite{Arutyunov:2004vx}. In order to test this assumption and to provide an independent check of these results we exploit the additional information coming from the scattering of mixed-flux dyonic giant magnons and plane waves to derive the semiclassical and 1-loop phases, as was done in the cases of $AdS_5\times S^5$ in \cite{Chen:2006gq,*Roiban:2006gs,Chen:2007vs} and for $AdS_3\times S^3$ in \cite{David:2010yg}.\\
\\
In string theory on $AdS_3\times S^3\times T^4$ with mixed 3-form flux dyonic giant magnons are classical soliton solutions moving on the $R\times S^3$ subspace with the dispersion relation \cite{Hoare:2013lja,Hofman:2006xt,Chen:2006gea}
\begin{align}
\varepsilon \equiv E-J_1 = \sqrt{(J\pm q\hh p)^2 +4\hh^2\qhat^2\sin^2\frac{p}{2}},\qquad\qhat=\sqrt{1-q^2},\qquad E, J_1\rightarrow \infty,\label{dispRel}
\end{align}
where $q\in(0,1)$ is the coefficient of the NS-NS flux, $(J_1,J)$ are two angular momenta and the world-sheet momentum $p$ is related to the effective kink-charge corresponding to the opening angle between the end-points of the string along a circle of $S^3$.\\
\\
The parameter $\hh$ denotes the string tension in the semiclassical limit, i.e. the 't Hooft coupling $\lambda$ is large with $\hh$ given by
\begin{align}
 \hh = \frac{\sqrt{\lambda}}{2\pi} = \frac{R^2}{2\pi\alpha'},
\end{align}
where $R$ is the curvature radius of $AdS_3$ and $S^3$. This relation could receive corrections in $1/\sqrt{\lambda}$ as in the case of $AdS_4\times CP^3$ \cite{Nishioka:2008gz,*Gaiotto:2008cg,*Gromov:2014eha}. In the action $\hh$ appears in the coefficient $\hh q/2$ of the WZ term for the NS-NS flux and as such the combination 
\begin{align}
 2\pi \hh q = q\sqrt{\lambda}
\end{align}
is the quantised WZ level.\\
\\
In the quantum theory soliton solutions are associated to asymptotic states. Their dispersion relation and S-matrix $S(p_1,p_2)= \exp(i\Theta(p_1,p_2))$ have the semiclassical expansion, i.e. $h\gg 1$,
\begin{align}
E(p) &= \hh E_{\text{cl}}(p) + \Delta E(p) + \mathcal{O}\Big(\frac{1}{\hh}\Big),\\
\Theta (p_1,p_2) &= \hh \Theta_{\text{cl}}(p_1,p_2) + \Delta \Theta(p_1,p_2) + \mathcal{O}\Big(\frac{1}{\hh}\Big).
\end{align}
For classical scattering, integrability implies that such soliton solutions experience only an overall time-delay, $\Delta T$, relative to free propagation. This time-delay is related to the leading order S-matrix by the WKB approximation \cite{Jackiw:1975im} 
\begin{align}
\Theta_{\text{cl}}(p_1,p_2) &= \frac{1}{\hh}\int_{E_{\text{th}}}^{E(p_1)}\dif E_1\, \Delta T (E_1,E_2;J_1,J_2) \label{thcl}
\end{align}
where $E_{\text{th}} = E_1|_{p_1=0}$. For dyonic giant magnon solitons the semiclassical limit corresponds to
\begin{align}
\hh\rightarrow \infty,\qquad \frac{J}{\hh}\,\,\mathrm{fixed},\qquad p\,\,\mathrm{fixed},
\end{align}
with associated bound states in the quantum theory such that their dispersion relation has the same form \eqref{dispRel} as the classical solution
\begin{align}
 E_{\text{cl}} = \frac{1}{\hh} \varepsilon = \sqrt{\Big(\frac{J}{\hh}\pm qp\Big)^2 +4\qhat^2\sin^2\frac{p}{2}}.
\end{align}
Further the 1-loop corrections were obtained for an integrable field theory in \cite{Chen:2007vs}
\begin{align}
\Delta E(p) &= \frac{1}{2\pi}\sum_{I=1}^{N_F}(-1)^{F_I}\int_{-\infty}^\infty \dif k\,\frac{\partial\delta_I(k;p)}{\partial k}\omega(k)\label{deltaE}\\
\Delta \Theta(p_1,p_2) &= \frac{1}{2\pi}\sum_{I=1}^{N_F}(-1)^{F_I}\int_{-\infty}^\infty \dif k\,\frac{\partial\delta_I(k;p_1)}{\partial k}\delta_I(k;p_2)\label{deltaTheta}.
\end{align}
where the $\delta_I(k;p)$ are the phase shifts for plane waves of momentum $k$ scattering off classical soliton solutions of momentum $p$. These plane waves have the dispersion relation $\omega(k)$ and they represent small fluctuations around the soliton solution labelled by the index $I=1,..,N_F$, with $(-1)^{F_I}=1$ for bosnonic and $(-1)^{F_I}=-1$ for fermionic fields.\\
\\
For an integrable theory the resulting bound-state S-matrix corresponds to the fusion of S-matrices for the scattering of elementary constituents \cite{Zamolodchikov1979253,*Karowski:1978ps}. This will allow us to obtain the elementary semiclassical and 1-loop phases from the above dyonic giant magnon bound-state picture. To evaluate the above expressions we first need to find the time-delay for dyonic giant magnon scattering and the phase shifts for plane wave scattering off dyonic giant magnons. Both can be obtained by constructing multi-soliton scattering solutions using the dressing method \cite{Spradlin:2006wk,*Kalousios:2006xy}.\\
\\
In section \ref{sec:dressing} we extend this method to the case of mixed flux and as a consitency check we reproduce the mixed-flux dyonic giant magnon by dressing the BMN solution. Applying the dressing method again we find the scattering solution for two dyonic giant magnons from which we extract the time-delay. In general such classical solutions on $R\times S^3$ are related to soliton solutions of the Complex sine-Gordon (CsG) model via Pohlmeyer reduction \cite{Pohlmeyer:1975nb} (for $q\ne 0$ also see \cite{Hoare:2013pma}). Scattering solutions in the CsG model can then be obtained using Bäcklund transformations and for $q=0$ the soliton for the scattering of two dyonic giant magnons is known explicitly. In appendix \ref{app:csg} we generalise this soliton to $q\ne 0$ and we find the time-delay from the CsG picture as an independent check for our result.\\
\\
In section \ref{sec:leadingOrder} we determine the bound-state S-matrix using the WKB formula \eqref{thcl}. The bound-state S-matrix includes contributions from the fusion of the elementary dressing (AFS) phase as well as the BDS factors and its equivalent for mixed excitation scattering. Eliminating the latter and using arguments relating to the functional form of the resulting AFS contribution we arrive at the dressing phase for elementary excitations.\\
\\
In section \ref{sec:1loop} we turn our attention to the 1-loop corrections \eqref{deltaE} and \eqref{deltaTheta}. Using the generalised dressing method we determine the bosonic phase shifts experienced by small fluctuations when scattering off dyonic giant magnons in $AdS_3\times S^3$. We do not consider directions along $T^4$ as we are interested in the massive sector only.\\
\\ 
We then establish the fermionic phase shifts as well as the explicit form of the dyonic giant magnon as a finite-gap solution by considering the connection between bosonic phase shifts and finite-gap equations. This also serves as an additional check for the finite-gap picture obtained in \cite{Babichenko:2014yaa}. Finally in section \ref{sec:1loopeval} we use this scattering data to evaluate the 1-loop correction to the bound-state S-matrix. Following similar arguments as in the leading order case we find the 1-loop elementary dressing phases which agree with \cite{Babichenko:2014yaa,Engelund:2013fja,*Bianchi:2014rfa} in the semiclssical limit ($\hh\rightarrow\infty, \hh p=\mathrm{fixed}$). We also find that the 1-loop energy shift \eqref{deltaE} vanishes.

\section{Dressing method for solutions on \texorpdfstring{$\R\times S^3$}{R x S3} with NS-NS flux}\label{sec:dressing}

In this section we extend the dressing method \cite{Zakharov:1973pp,*Zakharov:1980ty} for solutions on $\R\times S^3$ to the $q\ne 0$ case and applying this method to the BMN solution we reproduce the mixed-flux dyonic giant magnon.  In our analysis we closely follow \cite{Spradlin:2006wk,*Kalousios:2006xy}. We then apply the dressing method to the dyonic giant magnon to find the explicit scattering solution for two dyonic giant magnons from which we read off the time-delay.\\
\\
In conformal gauge the string action is equivalent to a principal chiral model with a Wess-Zumino term with a coefficient $q\in (0,1)$
\begin{align}\label{21}
S = -\frac{\hv}{2}\Big[\int\dif^2\sigma\,\tfrac{1}{2}\tr(\alg{J}_+\alg{J}_-)
- q\int\dif^3\sigma\,\tfrac{1}{3}\varepsilon^{abc}\tr(\alg{J}_a\alg{J}_b\alg{J}_c)\Big]\ ,
\qquad \alg{J}_a = g^{-1}\partial_a g\ ,
\end{align}
where $\hh=\frac{\sqrt{\lambda}}{2\pi}$ is the string tension, $g\in SU(2)$ and $\sigma^\pm = \frac12 (\tau
\pm \sigma)$,\ $\partial_\pm = \partial_\tau \pm \partial_\sigma$.
The equations of motion
\begin{align}
\partial_\pm \tilde{\alg{J}}_\mp \pm \frac{1}{2}[\tilde {\alg{J}}_+,\tilde {\alg{J}}_-] = 0,\qquad \tilde {\alg{J}}_\pm = (1 \pm q) {\alg{J}}_\pm\label{eoms}
\end{align}
can be rephrased as the compatibility condition of the Lax pair equations
\begin{align}
\partial_\pm \Psi(\sigma^+,\sigma^-;\lambda) = \Psi(\sigma^+,\sigma^-;\lambda)A_\pm^{(\lambda)}\label{linSys}
\end{align}
where $\lambda$ is the spectral parameter and
\begin{align}
A_\pm^{(\lambda)} = \frac{1}{1\pm \lambda}\tilde {\alg{J}}_\pm = \frac{1 \pm q}{1\pm \lambda} {\alg{J}}_\pm.
\end{align}
We shall also impose unitarity
\begin{align}
\Psi^\dagger (\bar\lambda)\Psi(\lambda) =1.\label{unitarity}
\end{align}
Given a solution $\Psi(\lambda)$ to the linear system \eqref{linSys} we can obtain a solution to the equations of motion \eqref{eoms} by taking $g=\Psi(q)$. Conversely given a solution $g$ to the equations of motion we can solve the linear system for $\Psi(\lambda)$ such that $\Psi(q)=g$.\\
\\
Starting with a given solution $\Psi$ we can perform a $\lambda$-dependent gauge transformation to obtain a new solution
\begin{align}
\Psi   &\rightarrow \Psi' = \Psi\chi\\
A_\pm  &\rightarrow A_\pm' = \chi^{-1} A_\pm \chi + \chi^{-1}\partial_\pm \chi.
\end{align}
Here we need to choose $\chi(\lambda)$ such that the transformed $SU(2)$ current
\begin{align}
{\alg{J}}_\pm' = \chi^{-1} {\alg{J}}_\pm \chi + \frac{1\pm \lambda}{1\pm q}\chi^{-1}\partial_\pm \chi
\end{align}
is independent of $\lambda$. This can be done by requiring that $\chi$ is a meromorphic function with $\chi \rightarrow 1$ as $\lambda \rightarrow \infty$. The simplest choice is then a single pole at $\lambda = \lambda_1$. The unitarity condition \eqref{unitarity} then requires
\begin{align}
\chi^\dagger (\bar\lambda)\chi(\lambda)=1
\end{align}
which fixes $\chi$ to be of the form
\begin{align}
\chi = 1 + \frac{\lambda_1-\bar\lambda_1}{\lambda-\lambda_1}P
\end{align}
where $P$ is a nilpotent hermitian operator, i.e. $P=P^2 = P^\dagger$. Finally it remains to choose $P$ such that ${\alg{J}}_\pm'$ has no poles at $\lambda=\lambda_1$. This is achieved by choosing $P$ such that its image is spanned by $\{\Psi^{-1}(\lambda_1)e_1,\Psi^{-1}(\lambda_1)e_2,...\}$ where $e_i$ are arbitrary constant vectors. For our purposes $P$ shall have rank 1 and is explicitly given by
\begin{align}
 P = \frac{\Psi^{-1}(\lambda_1)ee^\dagger \Psi(\bar\lambda_1)}{e^\dagger \Psi(\bar\lambda_1)\Psi^{-1}(\lambda_1)e}.
\end{align}
The resulting dressing factor $\chi$ has the determinant
\begin{align}
\det \chi = \frac{\lambda - \bar\lambda_1}{\lambda - \lambda_1}
\end{align}
and the dressed solution is given by
\begin{align}
g_{\mathrm{dressed}} = \sqrt{\frac{\lambda_1-q}{\bar\lambda_1-q}}\Psi(q)\chi(q)
\end{align}
where the normalisation factor ensures that $g\in SU(2)$.

\subsection{Scattering solution and time-delay}

Using this method let us now derive the $q\ne 0$ dyonic giant magnon solution by dressing up the BMN solution
\begin{align}
Z_1 = e^{it},\qquad Z_2 = 0.
\end{align}
Parametrising $SU(2)$ in terms of the embedding coordinates as
\begin{align}
g = \left(\begin{array}{cc}
Z_1 & Z_2 \\ -\cc{Z}_2 & \cc{Z}_1
\end{array}\right),\qquad \lvert Z_1\rvert^2 + \lvert Z_2 \rvert^2 = 1
\end{align}
we have for the BMN solution
\begin{align}
g = \left(\begin{array}{cc}
e^{i(\sigma^+ + \sigma^-)} & 0 \\ 0 & e^{-i(\sigma^+ + \sigma^-)}
\end{array}\right),\quad \Psi(\lambda) = \left(\begin{array}{cc}
e^{iZ(\lambda)} & 0 \\ 0 & e^{-iZ(\lambda)}
\end{array}\right),\\Z(\lambda) = \frac{1+q}{1+\lambda}\sigma^+ + \frac{1-q}{1-\lambda}\sigma^-.
\end{align}
Since the projection operator $P$ does not depend on the scale of $e$ we can parametrise $e$ as
\begin{align}
e = (c, 1/c),\quad c\in \mathbb{C}^*.
\end{align}
Furthermore $w$ only enters $P$ in
\begin{align}
\Psi^{-1}(\lambda_1)e = \left(\begin{array}{c}
c\, e^{iZ(\lambda_1)} \\  c^{-1} e^{-iZ(\lambda_1)}
\end{array}\right)
\end{align}
allowing us to absorb $c$ by shifting $Z(\lambda_1) \rightarrow Z(\lambda_1) + i \log c$ which corresponds to a shift in $(\sigma^+, \sigma^-)$. Therefore we can set $c=1$ without loss of generality. The projector $P$ is then given by
\begin{align}
P = \frac{1}{1+e^{2i(Z(\lambda_1)-Z(\bar\lambda_1))}}\left(\begin{array}{cc}
1 & e^{-2iZ(\bar\lambda_1)} \\ e^{2iZ(\lambda_1)} & e^{2i(Z(\lambda_1)-Z(\bar\lambda_1))}
\end{array}\right)
\end{align}
and the dressed solution takes the form
\begin{align}
 Z_1 = \frac{e^{it}}{\lvert \tilde \lambda_1\rvert} \frac{\tilde\lambda_1 e^{2iZ(\lambda_1)}+\tilde{\bar\lambda}_1 e^{2iZ(\bar\lambda_1)}}{e^{2iZ(\lambda_1)}+e^{2iZ(\bar\lambda_1)}},\qquad 
 Z_2 = \frac{e^{-it}}{\lvert \tilde \lambda_1\rvert} \frac{i(\tilde\lambda_1-\tilde{\bar\lambda}_1)}{e^{2iZ(\lambda_1)}+e^{2iZ(\bar\lambda_1)}}.\label{ZgmSol}
\end{align}
where $\tilde \lambda_1 = \lambda_1 - q$, $\tilde{\bar \lambda}_1 = \bar\lambda_1 -q$. This is the $q\ne 0$ dyonic giant magnon solution which we can see by parametrising $\lambda_1$ as
\begin{align}
 \lambda_1 = q + re^{ip/2}\label{rEq}
\end{align}
and introducing the world-sheet coordinates
\begin{align}
u = i(Z(\lambda_1)-Z(\bar\lambda_1)),\qquad v = t+qx-(Z(\lambda_1)+Z(\bar\lambda_1)).
\end{align}
Equivalently we can parametrise these coordinates in terms of a rapidity $\theta$ and a parameter $\rho$ as
\begin{align}
u = \cos\rho ({\cal X} + q {\cal T}),\qquad v = \sin\rho ({\cal T} + q {\cal X}),\label{uvcoords}\\
{\cal X} = x\cosh\theta- t\sinh\theta, \qquad {\cal T} = t\cosh\theta- x\sinh\theta.
\end{align}
These dyonic giant magnon soliton parameters $\theta$ and $\rho$ are related to $q$, $r$ and $p$ as
\begin{align}
\tanh\theta = 2\frac{q+r\cos\frac{p}{2}}{1+r^2 + q^2+2qr\cos\frac{p}{2}},\qquad \cot\rho = \frac{2r\sin\frac{p}{2}}{r^2-1+q^2+2qr\cos\frac{p}{2}}\\
\tan\frac{p}{2} = \frac{\cos\rho}{\sinh\theta-q\cosh\theta+q\sin\rho}
\end{align}
and we can solve these relations explicitly for $r$ giving
\begin{align}
r^2 = \frac{\cosh\theta+\sin\rho-2q\sinh\theta}{\cosh\theta-\sin\rho}+q^2.
\end{align}
In terms of the energy $E$, angular momentum $J$ and the world-sheet momentum $p$ this becomes
\begin{align}
 r=\frac{E \pm M_\mp}{2h\sin\frac{p}{2}},\qquad M_\pm = J \pm qhp
\end{align}
where the sign distinguishes left and right movers. The solution \eqref{ZgmSol} then takes the standard $q\ne 0$ dyonic giant magnon form
\begin{align}
Z_1 = e^{it}(\cos\frac{p}{2}+i\sin\frac{p}{2}\tanh u),\qquad Z_2 = \sin\frac{p}{2}e^{i(v-qx)}\,\mathrm{sech}\, u.
\end{align}
Applying the dressing method a second time gives the scattering solution of two magnons
\begin{align}
Z_1 &= \frac{e^{it}}{2\lvert \tilde \lambda_1 \rvert\lvert \tilde \lambda_2 \rvert}\frac{R+ \lvert \tilde\lambda_1\rvert^2\lambda_{1\bar 1}\lambda_{2\bar 2} e^{-i(v_1-v_2)}+\lvert\tilde\lambda_2\rvert^2 \lambda_{1\bar 1}\lambda_{2\bar 2}e^{i(v_1-v_2)}  }{\lambda_{12}\lambda_{\bar 1\bar 2}\cosh(u_1+u_2)+\lambda_{1\bar 2}\lambda_{\bar 1 2}\cosh(u_1-u_2)+\lambda_{1\bar 1} \lambda_{2\bar 2} \cos(v_1-v_2)}\\
Z_2 &= \frac{e^{-iqx}}{2\lvert \tilde \lambda_1 \rvert\lvert \tilde \lambda_2 \rvert}\frac{\lambda_{\bar 11}e^{iv_1}(\lambda_{\bar 1\bar 2}\lambda_{1\bar 2}\tilde\lambda_2e^{u_2}+\lambda_{12}\lambda_{\bar 1 2}\bar{\tilde\lambda}_2 e^{-u_2})+ \lambda_{\bar 22}e^{iv_2}(\lambda_{\bar 2\bar 1}\lambda_{2\bar 1}\tilde\lambda_1 e^{u_1}+\lambda_{21}\lambda_{\bar 21}\bar{\tilde\lambda}_1e^{-u_1})}{\lambda_{\bar 1\bar 2}\lambda_{12}\cosh(u_1+u_2)+\lambda_{\bar 1 2}\lambda_{1\bar 2}\cosh(u_1-u_2)+\lambda_{\bar 11}\lambda_{\bar 22}\cos(v_1-v_2)}\\
R &= \lambda_{12}\lambda_{\bar 1\bar 2} (\tilde\lambda_1\tilde\lambda_2 e^{u_1+u_2}+\bar{\tilde\lambda}_1\bar{\tilde\lambda}_2 e^{-u_1-u_2})+\lambda_{\bar 1 2}\lambda_{1\bar 2}(\tilde\lambda_1\bar{\tilde\lambda}_2 e^{u_1-u_2}+\bar{\tilde\lambda}_1\tilde\lambda_2 e^{-u_1+u_2})
\end{align}
where
\begin{align}
\lambda_{ij} = \lambda_i - \lambda_j,\qquad \lambda_{\bar i} = \bar\lambda_i.
\end{align}
In terms of the familiar coordinates \eqref{uvcoords} this solution takes the form
\begin{align}
Z_1 &= e^{it}\frac{[\cos(v_1-v_2)\cosh \rho + i \sin(v_1-v_2)\sinh \rho] \sin\frac{p_1}{2}\sin\frac{p_2}{2}\sech u_1\sech u_2 + R+iI}{\sin\frac{p_1}{2}\sin\frac{p_2}{2}[\cos(v_1-v_2)\sech u_1\sech u_2+\tanh u_1\tanh u_2]+\cos\frac{p_1}{2}\cos\frac{p_2}{2}-\cosh \rho}\\
Z_2 &= e^{-iqx}\frac{V_1+V_2}{\sin\frac{p_1}{2}\sin\frac{p_2}{2}[\cos(v_1-v_2)\sech u_1\sech u_2+\tanh u_1\tanh u_2]+\cos\frac{p_1}{2}\cos\frac{p_2}{2}-\cosh \rho}
\end{align}
\begin{align}
R &= \cos^2\frac{p_1}{2}+\cos^2\frac{p_2}{2}-\cosh \rho \cos\frac{p_1}{2}\cos\frac{p_2}{2}-1\\
I &= \tanh u_1 \sin\frac{p_1}{2}\Big(\cos\frac{p_1}{2}-\cosh \rho\cos\frac{p_2}{2}\Big) + \tanh u_2 \sin\frac{p_2}{2}\Big(\cos\frac{p_2}{2}-\cosh \rho\cos\frac{p_1}{2}\Big)\\
V_1 &= e^{iv_1}\sin\frac{p_1}{2}\sech u_1\Big[i\Big(\cosh \rho\cos\frac{p_2}{2}-\cos\frac{p_1}{2}\Big)+\sinh \rho\sin\frac{p_2}{2}\tanh u_2\Big]\\
V_2 &= e^{iv_2}\sin\frac{p_2}{2}\sech u_2\Big[i\Big(\cosh \rho\cos\frac{p_1}{2}-\cos\frac{p_2}{2}\Big)-\sinh \rho\sin\frac{p_1}{2}\tanh u_1\Big]\\
\rho &= \ln \frac{r_2}{r_1}.
\end{align}
We can extract the time delay by comparing the scattering solution to a freely propagating giant magnon as $t\rightarrow\pm\infty$. The velocity of a single free propagating giant magnon is given by
\begin{align}
v_s = \frac{1}{\hh}\frac{\dif E}{\dif p} = \frac{v-q}{1-qv} = \frac{\sinh\theta-q\cosh\theta}{\cosh\theta-q\sinh\theta}.
\end{align}
Taking $x=v_s t$ and $t\rightarrow +\infty$ the freely propagating soliton phase shifted by $\delta t_+$ has the asymptotic form
\begin{align}
Z_1^{(1)} &= e^{i(t-\delta t_+)} \Big(\cos\frac{p_1}{2}+i\sin\frac{p_1}{2}\tanh\big[\cos\rho_1 (\sinh\theta_1-q\cosh\theta_1) \delta t_+\big]\Big),& Z_1^{(2)} &= e^{i(t-\delta t_+ + p_2/2)}\\
Z_2^{(1)} &= e^{i(v_1-q v_{s1} t)} \sin\frac{p_1}{2}\sech \big[\cos\rho_1 (\sinh\theta_1-q\cosh\theta_1) \delta t_+\big], & Z_2^{(2)} &= 0
\end{align}
whereas the scattering solution asymptotes to
\begin{align}
Z_1^{(s)} &= e^{it}\,\frac{e^{i\frac{p_2}{2}}(\cos\frac{p_2}{2}-\cos\frac{p_1}{2}\cosh \rho)-\sin^2\frac{p_1}{2}}{\cos\frac{p_1}{2}\cos\frac{p_2}{2}-\cosh \rho} &\Rightarrow
\lvert Z_1^{(s)}\rvert^2 &= \cos^2\frac{p_1}{2}+\frac{\sin^4\frac{p_1}{2}\cos^2\frac{p_2}{2}}{(\cosh \rho-\cos\frac{p_1}{2}\cos\frac{p_2}{2})^2}.
\end{align}
In the COM frame ($\sum \cos\rho_i(\sinh \theta_i-q\cosh\theta_i)=0$) both solitons experience the same time delay giving in total
\begin{align}
\Delta T = 2\delta t_+ = \frac{1}{(\sinh\theta_1-q\cosh\theta_1)\cos\rho_1}\Big\lvert\ln \frac{\cosh(\theta_1-\theta_2)-\cos(\rho_1-\rho_2)}{\cosh(\theta_1-\theta_2)+\cos(\rho_1+\rho_2)}\Big\rvert.\label{timeDelay}
\end{align}
For $q=0$ this reduces to the standard expression for the time-delay due to the collison of two solitons in the CsG model \cite{Dorey:1994mg}.

\section{Semiclassical bound-state S-matrix}\label{sec:leadingOrder}

In order to evaluate the semiclassical bound-state S-matrix
\begin{align}
\Theta (p_1,p_2) &= \int_0^{p_1} \dif p_1' \frac{\dif E}{\dif p}(p_1')\,\Delta T (p_1',p_2)\label{thcl2}
\end{align}
let us rewrite the above time-delay expression \eqref{timeDelay} in terms of the momentum $p_i$, angular momentum $J_i$ and energy $E_i$ of the associated individual dyonic giant magnons. In terms of the soliton parameters $\theta_i$ and $\rho_i$ these charges are given by \cite{Hoare:2013lja}\footnote{For the relation between $(E_i,J_i)$ and the currents $\alg{J}_\pm$ also see section 2.2 of \cite{Hoare:2013lja}.}
\begin{align}
 E_l = \sqrt{M^2+4\hh^2(1-q^2)\sin^2 \frac{p}{2}},\quad M_l \equiv  J + ql\hh p,\quad 
 M_l =  - 2l\hh\sin\frac{p}{2}(\tan\rho \sin\frac{p}{2} - q\cos\frac{p}{2})\label{emeqs}\\
 \sinh\theta = \frac{m + q\sqrt{m^2+1}}{\sqrt{1-q^2}},\quad \cosh\theta = \frac{qm +\sqrt{m^2+1}}{\sqrt{1-q^2}},\quad m = \frac{1}{\sqrt{1-q^2}}\Big(\frac{\cos\rho}{\tan\frac{p}{2}}-q\sin\rho\Big)\\
 \cosh(\theta_1-\theta_2) =\sqrt{ m_1^2+1}\sqrt{ m_2^2+1} - m_1 m_2
\end{align}
where $l=\pm 1$ represents left and right movers which are related by sending $J\rightarrow -J$. The time delay becomes
\begin{align}
\Delta T_{l_1l_2} &= C_{l_1} L_{l_1l_2} & C_{l_1} &= \frac{4\hh^2\sin^4\frac{p_1}{2}+(M_1 - q\hh l_1\sin p_1)^2}{2\sin^2\frac{p_1}{2}(2\hh^2\qhat^2\cos\frac{p_1}{2}\sin\frac{p_1}{2} + q\hh l_1 M_1)}\\
         &      & L_{l_1l_2} &= \Big\lvert\ln \frac{E_1E_2- \hh^2\qhat^2\sin p_1\sin p_2-4\hh^2\qhat^2\sin^2\frac{p_1}{2}\sin^2\frac{p_2}{2} - l_1l_2M_1 M_2}{E_1E_2- \hh^2\qhat^2\sin p_1\sin p_2+4\hh^2\qhat^2\sin^2\frac{p_1}{2}\sin^2\frac{p_2}{2} - l_1l_2M_1M_2}\Big\rvert.
\end{align}
We can further simplify \eqref{thcl2} by transforming the integral over energy into a contour integral over the dyonic spectral parameters
\begin{align}
 X^\pm(p) = \frac{e^{\pm ip/2}}{2\hh\qhat\sin\frac{p}{2}}\big[E(p) + M(p)\big],\qquad \qhat=\sqrt{1-q^2}
\end{align}
which for $J=1$ reduce to the usual spectral parameters $x^\pm$ corresponding to fundamental magnon excitations and which satisfy
\begin{align}
X^+ &+ \frac{1}{X^+} - \Big(X^- + \frac{1}{X^-} \Big) = \frac{2i}{\hh\qhat}M,\quad \frac{X^+}{X^-} = e^{ip}\label{shortcond}\\
X^+ &- \frac{1}{X^+} - \Big(X^- - \frac{1}{X^-} \Big) = \frac{2i}{\hh\qhat}E.
\end{align}
In terms of these variables the time delay can be written as
\begin{align}
C_{l_1} &= 
i\frac{(1+X^+X^-)^2-(\qhat(X^++X^-) - l_1 q(1-X^+X^-))^2}{\qhat(X^+-X^-)(\qhat(X^++X^-) - l_1 q(1-X^+X^-))},\\
L_{l_1l_2} &= -\begin{cases}
\ln \frac{X^--Y^-}{X^+-Y^-}\frac{X^+-Y^+}{X^--Y^+}, & l_1l_2 = 1\\
\ln \frac{\Big(1-\frac{1}{X^+ Y^-}\Big)\Big(1-\frac{1}{X^- Y^+}\Big)}{\Big(1-\frac{1}{X^+ Y^+}\Big)\Big(1-\frac{1}{X^- Y^-}\Big)}, & l_1l_2 = -1.
\end{cases}
\end{align}
Noting the identity
\begin{align}
\frac{\dif E_1}{\dif p_1} C_{l_1}     &= \frac{\dif X^\pm}{\dif p_1} \hh\qhat \frac{1-(X^\pm)^2+2l_1\frac{q}{\qhat}X^\pm}{(X^\pm)^2}
\end{align}
we can split the integral in \eqref{thcl2} into two separate integrals over $X^+$ and $X^-$ respectively giving
\begin{align}
\frac{1}{\hh\qhat}\Theta (X,Y) &= I(X^+,Y^+) - I(X^+,Y^-)  + I(X^-,Y^-) - I(X^-,Y^+)\\
I(X^\pm,Y)  &= \int_{X^\pm(0)}^{X^\pm(p_1)} \dif z\, \frac{1-z^2+2\frac{ql_1}{\qhat}z}{z^2}\begin{cases}
-\ln (z-Y), & l_1l_2 = 1\\
\ln \Big(1-\frac{1}{z Y}\Big), & l_1l_2 = -1
\end{cases}
\end{align}
where for $l_1l_2=1$ the integral evaluates to
\begin{align}
   I(X,Y) &= -X+\frac{1}{Y}\ln X - \Big[Y + \frac{1}{Y}-\Big(X+\frac{1}{X}\Big)+2l_1\frac{q}{\qhat}\ln\frac{X}{Y}\Big]\ln(X-Y)\\
          &  -2l_1\frac{q}{\qhat}\mathrm{Li}_2\Big(1-\frac{X}{Y}\Big) - I\Big\rvert_{p_1=0}
\end{align}
and for $l_1l_2=-1$ we have
\begin{align}
 \bar I(X,Y) &= \frac{1}{X}-\frac{\ln Y}{Y}+\frac{\ln X}{Y} +  \Big[Y+\frac{1}{Y}-(X+\frac{1}{X})\Big]\ln\Big(1-\frac{1}{XY}\Big)\\
             &  +2l_1\frac{q}{\qhat}\Li_2\Big(\frac{1}{XY}\Big)- I\Big\rvert_{p_1=0}.
\end{align}
The last term $I\rvert_{p_1=0}$ in these integrals corresponds to an infinite contribution from the integral at $p_1=0$. These infinite contributions arise from the way we chose to split the integral and they cancel in the sum $I(X^+,Y)-I(X^-,Y)$.\\
\\
We can therefore write the bound-state scattering matrix for $l_1l_2=1$ as
\begin{align}
\Theta (X,Y)      &= \hh\qhat\Big[     K(X^+,Y^+) +      K(X^-,Y^-) -      K(X^+,Y^-)  -      K(X^-,Y^+)\Big] + p_1(E_2-J_2)\\
K(X,Y) &= -\Big[Y + \frac{1}{Y}-\Big(X+\frac{1}{X}\Big)+2l_1\frac{q}{\qhat}\ln\frac{X}{Y}\Big]\ln(X-Y) - 2l_1\frac{q}{\qhat}\mathrm{Li}_2\Big(1-\frac{X}{Y}\Big) +l_1\frac{q}{\qhat}\ln X\ln Y
\end{align}
and for $l_1l_2=-1$ as
\begin{align}
\bar \Theta (X,Y) &= \hh\qhat\Big[\bar K(X^+,Y^+) + \bar K(X^-,Y^-) - \bar K(X^+,Y^-)  - \bar K(X^-,Y^+)\Big] + p_1(E_2-J_2)\\
\bar K(X,Y) &= \Big[Y+\frac{1}{Y}-(X+\frac{1}{X})\Big]\ln\Big(1-\frac{1}{XY}\Big)+2l_1\frac{q}{\qhat}\Li_2\Big(\frac{1}{XY}\Big) -l_1\frac{q}{\qhat}\ln X\ln Y \label{bSmat}.
\end{align}
Here the terms proportional to $p_1$ are gauge terms originating from the choice of conformal gauge (for details see \cite{Hofman:2006xt}). The remaining part of the S-matrix contains another such contribution from the terms of $\ln X\ln Y$. However the gauge term is uniquely fixed by imposing that the S-matrix without the gauge term is antisymmetric under the exchange $X \leftrightarrow Y$ and $l_1\leftrightarrow l_2$ as required by unitarity.

\subsection{Dressing phases for elementary excitations}\label{sec:afsphases}

In order to obtain the elementary dressing phases we need to eliminate the bound-state contributions in the S-matrix coming from the BDS factor for same-type excitations and the equivalent factor for mixed-type excitations \cite{Lloyd:2014bsa}
\begin{align}
s_{BDS}(x,y) = \frac{x^+-y^-}{x^--y^+}\frac{1-\frac{1}{x^+y^-}}{1-\frac{1}{x^-y^+}},\qquad
& s_{\mathrm{mix}}(x,y) = \frac{1-\frac{1}{x^-y^+}}{1-\frac{1}{x^-y^-}}\frac{1-\frac{1}{x^+y^-}}{1-\frac{1}{x^+y^+}}.\label{rfact}
\end{align}
Integrability allows us to fuse together these elementary factors into their bound-state contributions
\begin{align}
S = \prod_{j_1=1}^{J_1}\prod_{j_2=1}^{J_2} s(x^+_{j_1},x^-_{j_1};y^+_{j_2},y^-_{j_2})\label{boundstateBDS}
\end{align}
where $x_{j_1}$, $y_{j_2}$ are the spectral parameters of the constituents satisfying the shortening conditions
\begin{align}
x^+_{j_1} + \frac{1}{x^+_{j_1}}-2l_X\frac{q}{\qhat}\ln x^+_{j_1} -\Big(x^-_{j_1}+\frac{1}{x^-_{j_1}}-2l_X\frac{q}{\qhat}\ln x^-_{j_1}\Big) &= \frac{2i}{\hh\qhat}, & l_X&=\pm 1, & j_1=1,..,J_1-1\\
y^+_{j_2} + \frac{1}{y^+_{j_2}}-2l_Y\frac{q}{\qhat}\ln y^+_{j_2} -\Big(y^-_{j_2}+\frac{1}{y^-_{j_2}}-2l_Y\frac{q}{\qhat}\ln y^-_{j_2}\Big) &= \frac{2i}{\hh\qhat}, & l_Y&=\pm 1, & j_2=1,..,J_2-1
\end{align}
and the pole conditions for the formation of the $J_1$ and $J_2$ bound states
\begin{align}
x^-_{j_1} &= x^+_{j_1+1},\qquad j_1=1,..,J_1-1\\
y^-_{j_2} &= y^+_{j_2+1},\qquad j_2=1,..,J_2-1.
\end{align}
The bound-state spectral parameters are then identified as
\begin{align}
X^+ &= x^+_1, & X^- &= x^-_{J_1}, &
Y^+ &= y^+_1, & Y^- &= y^-_{J_2}.
\end{align}
Let us first compute the semiclassical bound-state contribution for the BDS factor. Splitting off the $j_1=1$, $j_2=1$, $j_1=J_1$ and $j_2=J_2$ terms in the product \eqref{boundstateBDS} we get
\begin{align}
     S_{BDS} = & \exp(i\Theta_{BDS})\\
i\Theta_{BDS} = & \ln \frac{X^+-Y^-}{X^--Y^+}\frac{1-\frac{1}{X^+Y^-}}{1-\frac{1}{X^-Y^+}}
  + \sum_{j_2=1}^{J_2-1}\ln\frac{X^+-y^-_{j_2}}{X^--y^+_{j_2}}\frac{1-\frac{1}{X^+y^-_{j_2}}}{1-\frac{1}{X^-y^+_{j_2}}}
  + \sum_{j_1=2}^{J_1}\ln\frac{x_{j_1}^+-Y^-}{x_{j_1}^--Y^+}\frac{1-\frac{1}{x_{j_1}^+Y^-}}{1-\frac{1}{x_{j_1}^-Y^+}}.
\end{align}
Taking the semiclassical dyonic giant magnon limit $\hh\rightarrow\infty, J\sim \hh$ the sums transform into integrals giving
\begin{align}
i\Theta_{BDS} = & 
    \hh\int_0^{J_2/\hh}\dif j\, \ln\frac{X^+-y^-(j)}{X^--y^+(j)}\frac{1-\frac{1}{X^+y^-(j)}}{1-\frac{1}{X^-y^+(j)}}
  + \hh\int_0^{J_1/\hh}\dif j\, \ln\frac{x^+(j)-Y^-}{x^-(j)-Y^+}\frac{1-\frac{1}{x^+(j)Y^-}}{1-\frac{1}{x^-(j)Y^+}} + \mathcal{O}(1).\notag
\end{align}
In the semiclassical limit the shortening condition of the constituents and the pole condition combined give
\begin{align}
x^-(j) + \frac{1}{x^-(j)} - 2l_1\frac{q}{\qhat}\ln x^-(j) &= X^+ + \frac{1}{X^+} - 2l_1\frac{q}{\qhat}\ln X^+ -\frac{2i}{\qhat}j\\
y^-(j) + \frac{1}{y^-(j)} - 2l_2\frac{q}{\qhat}\ln y^-(j) &= Y^+ + \frac{1}{Y^+} - 2l_2\frac{q}{\qhat}\ln Y^+ -\frac{2i}{\qhat}j
\end{align}
where $j\sim \hh$ has been rescaled by $\hh$. This allows us to rewrite the BDS contribution in terms of contour integrals over the spectral parameters as
\begin{align}
\Theta_{BDS} &= 
    \frac{\hh\qhat}{2}\int_{Y^+}^{Y^-}\dif z\, \Big(1-\frac{1}{z^2}-2l_2\frac{q}{\qhat}\frac{1}{z}\Big)\ln\frac{X^+-z}{X^--z}\frac{1-\frac{1}{X^+z}}{1-\frac{1}{X^-z}}\\ 
    &+\frac{\hh\qhat}{2}\int_{X^+}^{X^-}\dif z\, \Big(1-\frac{1}{z^2}-2l_1\frac{q}{\qhat}\frac{1}{z}\Big)\ln\frac{z-Y^-}{z-Y^+}\frac{1-\frac{1}{zY^-}}{1-\frac{1}{zY^+}} 
  + \mathcal{O}(1).
\end{align}
Performing these integrals we obtain the bound-state BDS contribution
\begin{align}
\Theta_{BDS} &= \hh\qhat\big[\hat k(X^-,Y^-)- \hat k(X^+,Y^-)  - \hat k(X^-,Y^+) + \hat k(X^+,Y^+)\big] + \mathcal{O}(1)\\
\hat k(X,Y)  &= \Big[X+\frac{1}{X}-\Big(Y+\frac{1}{Y}\Big)\Big]\ln\Big[(X-Y)\Big(1-\frac{1}{XY}\Big)\Big]\\& -2l_1\frac{q}{\qhat}\ln\frac{X}{Y}\ln(X-Y)-l_1\frac{q}{\qhat}\Big(2\Li_2\Big(1-\frac{X}{Y}\Big)-\ln X\ln Y\Big).
\end{align}
For $q=0$ this agrees with the BDS contribution in the case of $AdS_5\times S^5$ given in \cite{Chen:2006gq} (see equation (34)). Performing this fusion procedure on the factor \eqref{rfact} for mixed-type magnon scattering one finds that the integrals cancel at the linear order in $\hh$, i.e.
\begin{align}
 \Theta_{\mathrm{mix}} &\sim \mathcal{O}(1).
\end{align}
Subtracting these contributions from the bound-state S-matrix \eqref{bSmat} we are left with
\begin{align}
\Theta_{AFS}(X,Y) &= \hh\qhat \Big[\chi(X^+,Y^+) -\chi(X^+,Y^-)-\chi(X^-,Y^+)+\chi(X^-,Y^-)\Big]\label{lo1}\\
\bar\Theta(X,Y) &= \hh\qhat \Big[\bar\chi(X^+, Y^+) -\bar\chi(X^+, Y^-)-\bar\chi(X^-, Y^+)+\bar\chi(X^-, Y^-)\Big]\\
\chi(x,y)     &= \Big(y + \frac{1}{y} - x - \frac{1}{x}\Big)\ln \Big(1-\frac{1}{xy}\Big)\\
\bar\chi(x,y) &= \Big(y + \frac{1}{y} - x - \frac{1}{x}\Big)\ln \Big(1-\frac{1}{xy}\Big) +l_1 \frac{q}{\sqrt{1-q^2}}\Big(2\mathrm{Li}_2\Big(\frac{1}{xy}\Big)-\ln x \ln y\Big)\label{lo2}
\end{align}
where $\Theta_{AFS}$ is the contribution for same-type magnons ($X$, $Y$ with $l_X=l_Y$) and $\bar\Theta$ is the contribution for mixed-type magnons ($X$, $Y$ with $l_X=-l_Y$).\\
\\
These bound-state AFS and mixed scattering contributions are related to the tree-level elementary dressing phases by the fusion procedure. However, since both the elementary dressing phases and the associated bound-state contributions are of the same leading order in $\hh$, the fusion procedure results in the same functional form for the bound-state result as for the elementary phases\footnote{This is because any sum of elementary factors contributes an order of $\mathcal{O}(\hh)$. However all such contributions must cancel for the bound-state and elementary phase to be of the same leading order $\mathcal{O}(\hh)$. For $q=0$ this is apparent from the particular form of the elementary dressing phase
\begin{align}
\theta(x,y) = f(x^+,y^+) - f(x^+,y^-) - f(x^-,y^+) + f(x^-,y^-)\label{elemForm}
\end{align}
which fused into the bound-state contribution $\Theta$ has the same functional form
\begin{align}
\Theta(X,Y) = f(X^+,Y^+) - f(X^+,Y^-) - f(X^-,Y^+) + f(X^-,Y^-).
\end{align}}.
Therefore we can deduce that the elementary dressing phase must have the functional form of the associated bound-state contribution \eqref{lo1}-\eqref{lo2} in agreement with the prediction in \cite{Babichenko:2014yaa}.
\section{One-loop corrections}\label{sec:1loop}

In this part of the paper we will determine the 1-loop corrections to the dispersion relation and the soliton S-matrix \eqref{deltaE}-\eqref{deltaTheta}. The involved phase shifts vanish for fluctuations on the $AdS_3$ and $T^4$ parts since we are only considering the dyonic giant magnon which moves in the $R\times S^3$ subspace of $AdS_3\times S^3\times T^4$. In order to find the phase shifts for fluctuations on $S^3$ we will use the dressing method which allows us to obtain multi-soliton scattering solutions. Identifying the limit in which one of these dyonic giant magnon solitons reduces to a plane wave we obtain a solution for a plane wave scattering off multiple dyonic giant magnons. From the asymptotic behaviour of this solution at $x\rightarrow \pm \infty$ we then find the associated bosonic phase shifts.\\ 
\\
The dressing method only covers bosonic phase shifts, however in the formulation of classical solutions in terms of the finite-gap picture fermionic and bosonic fluctuations are closely related. Exploiting this relation we will see in section \ref{sec:fermionic} how the information coming from the bosonic phase shifts is already sufficient to determine the fermionic phase shifts as well as the exact form of the dyonic giant magnon solution in the finite-gap picture. %
%
%
In section \ref{sec:1loopeval} we use this semiclassical scattering data to evaluate the 1-loop bound-state corrections \eqref{deltaE}-\eqref{deltaTheta} and we deduce the 1-loop corrections for the scattering of elementary excitations.
%

\subsection{Phase shifts for bosonic fluctuations}\label{sec:bosonic}


Let us first relate the plane wave solutions to the dyonic giant magnon. In order to obtain a plane wave solution to the linearised equations of motion we take the semiclassical limit 
\begin{align}
h\rightarrow \infty,\qquad k\equiv hp\,\,\mathrm{fixed},\qquad J\,\,\mathrm{fixed}.\label{planewaveLimit}
\end{align}
In this plane wave limit the spectral parameters have the expansion
\begin{align}
X^\pm &\sim w + \mathcal{O}\Big(\frac{1}{\hh}\Big), \qquad \ww=\frac{r}{\qhat},\quad 
     r = \frac{1}{k}\big[J+ qlk+\varepsilon(k)\big]\label{planewavereq} 
\end{align}
with the dispersion relation
\begin{align}
\varepsilon(k) = \sqrt{(k+qlJ)^2+\qhat^2J^2}.
\end{align}
In the large momentum limit the spectral parameter reduces to
\begin{align}
w_l \rightarrow s^{\pm l},\qquad k\rightarrow \pm \infty,\qquad s=\frac{\sqrt{1+q}}{\sqrt{1-q}}.
\end{align}
%
Expanding the dyonic giant magnon solution in the limit \eqref{planewaveLimit} we obtain a plane wave
\begin{align}
\delta Z_1 &= Z_1 - Z_1^{BMN} \sim 0\\
\delta Z_2 &= Z_2 - Z_2^{BMN} \sim \sin\frac{p}{2} e^{i(v-qx)} = \sin\frac{p}{2} \exp\Big(-i\frac{\varepsilon(k)t-kx}{Jl}\Big)
\end{align}
Since we are interested in the phase shifts for an elementary plane wave we take from now on $J=1$. Let us further parametrise its frequency $\omega$ and momentum $k$ in terms of the spectral parameter $\ww$ as
\begin{align}
\omega &= \sqrt{(k+ql)^2+\qhat^2}=\frac{1-q^2+r^2}{(r+ 1-lq)(r-(1+lq))} = \frac{\qhat(1+\ww_l^2)}{(\ww_l+s^{-l})(\ww_l-s^l)},\\
     k &= \frac{2r}{(r+ 1-lq)(r-(1+lq))} = \frac{2\ww_l}{\qhat(\ww_l+s^{-l})(\ww_l-s^l)}\label{planewaveRel}.
\end{align}
This dispersion relation corresponds to two separate plane waves with $l=\pm 1$ representing left or right movers. In the following derivation of the phase shifts we will only consider $l=-1$ plane waves and dyonic giant magnons for simplicity. In order to generalise these to arbitrary combinations of left and right moving plane waves and dyonic giant magnons we can simply send $J\rightarrow - J$. In terms of the dyonic giant magnon spectral parameters $X^\pm_l$ this corresponds to the transformation
\begin{align}
X^\pm_- \rightarrow \frac{1}{X^\mp_+}\label{Xl2r}
\end{align}
For plane waves we have $x^+ \sim x^- \sim \ww$ and this transformation reduces to
\begin{align}
\ww_- \rightarrow \frac{1}{\ww_+}.\label{wwl2r}
\end{align}
Using the dressing method we can now calculate the phase shift of a plane wave scattering off an $N$-soliton solution. For this we construct the asymptotic form of the $N$-soliton solution and take the plane wave limit in the $N+1$ dressing step. The recursive dressing relation is
\begin{align}
g_N = \sqrt{\frac{\lambda_N-q}{\bar\lambda_N-q}}\Psi_{N-1}(q)\Big(1+\frac{\lambda_N-\bar\lambda_N}{q-\lambda_N}P\Big).
\end{align}
Identifying
\begin{align}
\lambda_N = q + r_N e^{ip_N/2},\qquad \bar\lambda_N = q +r_N e^{-ip_N/2},\qquad \Psi_N(q) = g_N
\end{align}
we obtain
\begin{align}
g_N = g_{N-1}\Big(e^{ip_N/2}-2i\sin\frac{p_N}{2}P_{N-1}\Big).
\end{align}
Expanding this expression in the plane wave limit (in small $\eta=x^+-x^-=2ir \sin\frac{p}{2}$) gives the phase shifts
\begin{align}
\delta g = g_{N+1} - g_N &= g_{N}\Big(e^{ip/2}-2i\sin\frac{p}{2}P_{N}-1\Big)
\sim i\sin\frac{p}{2}\Big(g_N(1-2P_{N+1})\rvert_{\eta=0,x\rightarrow\pm\infty}\Big).
\end{align}
We now need to determine the $N$-soliton solution $g_N$ and the projector $P_N$ in the plane wave limit using the dressing transformation 
\begin{align}
\Psi_N(\lambda) = \Psi_{N-1}(\lambda)\chi_N(\lambda),\qquad \chi_N(\lambda)= 1+\frac{\lambda_N-\bar\lambda_N}{\lambda-\lambda_N}P_{N}.\label{drtr}
\end{align}
First let us show that the asymptotic form of the projector (without taking the plane wave limit) is
\begin{align}
P_i\lvert_{x\rightarrow +\infty} = \left(\begin{array}{cc}
0 & 0 \\ 0 & 1
\end{array}\right),\qquad 
P_i\lvert_{x\rightarrow -\infty} = \left(\begin{array}{cc}
1 & 0 \\ 0 & 0
\end{array}\right).\label{Pasymp}
\end{align}
We can establish this by induction. Starting with the projector $P_1$ for the giant magnon
\begin{align}
P_1 = \frac{1}{2}\sech u\left(\begin{array}{cc}
e^{-u} & e^{i(v-t-qx)} \\ e^{-i(v-t-qx)} & e^u
\end{array}\right)
\end{align}
we obtain the above asymptotic behaviour as $u\rightarrow\pm\infty$. Assuming this also holds for all the projectors up to $P_N$ we can construct the next projector $P_{N+1}$ by applying the dressing transformation using the above asymptotic form. The dressing transformation \eqref{drtr} becomes 
\begin{align}
\Psi_N &= e^{iP/2}\Psi_0\prod_{k=1}^N\chi_k(\lambda), & P =\sum_{k=1}^N p_k,&\qquad A(\lambda) = \prod_{k=1}^N\frac{\lambda-\bar\lambda_k}{\lambda-\lambda_k},\\
\chi_k(\lambda)\rvert_{x\rightarrow +\infty} &= \left(\begin{array}{cc}
1 & 0 \\ 0 & \frac{\lambda-\bar\lambda_k}{\lambda-\lambda_k}
\end{array}\right), & 
\Psi_N\rvert_{x\rightarrow+\infty} &= e^{iP/2}\left(\begin{array}{cc}
e^{iZ(\lambda)} & 0 \\ 0 & e^{-iZ(\lambda)}A(\lambda)
\end{array}\right),\\
\chi_k(\lambda)\rvert_{x\rightarrow -\infty} &= \left(\begin{array}{cc}
\frac{\lambda-\bar\lambda_k}{\lambda-\lambda_k} & 0 \\ 0 & 1
\end{array}\right), & 
\Psi_N\rvert_{x\rightarrow-\infty} &= e^{iP/2}\left(\begin{array}{cc}
e^{iZ(\lambda)}A(\lambda) & 0 \\ 0 & e^{-iZ(\lambda)}
\end{array}\right)\label{Nsol}
\end{align}
and using
\begin{align}
P_N = \frac{[\Psi_{N-1}(\bar\lambda_N)]^\dagger ee^\dagger \Psi_{N-1}(\bar\lambda_N)}{e^\dagger [\Psi_{N-1}(\bar\lambda_N)]^\dagger \Psi_{N-1}(\bar\lambda_N)e},\qquad e=(1,1)
\end{align}
we find the projector
\begin{align}
P_{N+1}\rvert_{x\rightarrow+\infty} &= \frac{1}{e^{-u}+e^u\lvert\Pi\rvert^2}\left(\begin{array}{cc}
e^{-u} & e^{i(v-t-qx)}\Pi \\
e^{-i(v-t-qx)}\bar\Pi & e^u \lvert\Pi\rvert^2
\end{array}\right)\rvert_{x\rightarrow+\infty},\qquad \Pi \equiv \prod_{k=1}^N\frac{\lambda_{N+1}-\bar\lambda_k}{\lambda_{N+1}-\lambda_k}\\
P_{N+1}\rvert_{x\rightarrow-\infty} &= \frac{1}{e^{-u}\lvert\Pi\rvert^2+e^u}\left(\begin{array}{cc}
e^{-u}\lvert\Pi\rvert^2 & e^{i(v-t-qx)}\bar\Pi \\
e^{-i(v-t-qx)}\Pi & e^u
\end{array}\right)\rvert_{x\rightarrow-\infty}.
\end{align}
Taking $u\rightarrow\pm\infty$ we arrive at the asymptotic behaviour \eqref{Pasymp} as required. However in the plane wave limit we have $\bar\lambda_{N+1}=\lambda_{N+1} = q + r + \mathcal{O}(1/\hh)$. Therefore by definition $u\equiv 0$ and $\bar\Pi=1/\Pi$ giving the projector
\begin{align}
 P_{N+1}\rvert_{x\rightarrow+\infty} &\sim \frac{1}{2}\left(\begin{array}{cc}
1 & e^{i(v-t-qx)}\Pi \\
e^{-i(v-t-qx)}\bar\Pi & 1
\end{array}\right),\\
P_{N+1}\rvert_{x\rightarrow-\infty} &\sim \frac{1}{2}\left(\begin{array}{cc}
1 & e^{i(v-t-qx)}\bar\Pi \\
e^{-i(v-t-qx)}\Pi & 1
\end{array}\right).
\end{align}
Finally the $N$-soliton solution as obtained from \eqref{Nsol} takes the form
\begin{align}
g_N\rvert_{x\rightarrow+\infty} = \left(\begin{array}{cc}
e^{i(t+P/2)} & 0 \\
0 & e^{-i(t+P/2)}
\end{array}\right),\qquad
g_N\rvert_{x\rightarrow-\infty} = \left(\begin{array}{cc}
e^{i(t-P/2)} & 0 \\
0 & e^{-i(t-P/2)}
\end{array}\right)
\end{align}
giving
\begin{align}
\delta g\rvert_{x\rightarrow\pm\infty} &\sim -i\sin\frac{p}{2}\left(\begin{array}{cc}
0 & e^{\pm iP/2}e^{i(v-qx)}\Pi^{\pm 1} \\
e^{\mp iP/2}e^{-i(v-qx)}\Pi^{\mp 1} & 0
\end{array}\right)
%
\end{align}
or in terms of the coordinates
\begin{align}
\delta Z_1\rvert_{x\rightarrow\pm\infty} = 0,\qquad 
\delta Z_2\rvert_{x\rightarrow\pm\infty} = -i \sin\frac{p}{2}e^{\pm iP/2}e^{i(v-qx)}\Pi^{\pm 1}\\
\delta \bar Z_1\rvert_{x\rightarrow\pm\infty} = 0,\qquad 
\delta \bar Z_2\rvert_{x\rightarrow\pm\infty} = i\sin\frac{p}{2}e^{\mp iP/2}e^{-i(v-qx)}\Pi^{\mp 1}.
\end{align}
Note that $v-qx = \omega t - kx$ such that these are plane wave solutions of frequency $\omega$ and momentum $k$ given by \eqref{planewaveRel}. For the conjugate fields we have $\omega \rightarrow -\omega$ and $k\rightarrow -k$ corresponding to $w_l\rightarrow 1/w_{-l}$. We read off the phase shifts
\begin{align}
\delta_{Z_1} &= 0\\
\delta_{Z_2} &\equiv i \ln \delta Z_2\rvert_{x\rightarrow+\infty} - i \ln \delta Z_2\rvert_{x\rightarrow-\infty} = -P + 2i\ln \Pi = -P - 2i\sum_{k=1}^N\ln\frac{\ww_--{X^+_-}_k}{\ww_--{X^-_-}_k}\\
\delta_{\bar Z_1} &= 0\\
\delta_{\bar Z_2} &= P - 2i\ln \Pi = P + 2i\sum_{k=1}^N\ln\frac{\frac{1}{\ww_+}-{X^+_-}_k}{\frac{1}{\ww_+}-{X^-_-}_k}
\end{align}
where we used \eqref{rEq} to write these expressions in terms of the spectral parameters ${X^\pm_\pm}_k$ with the lower sign denoting left or right movers. Finally we can obtain phase shifts for any combination of left and right moving plane waves and dyonic giant magnons using the transformations \eqref{Xl2r}, \eqref{wwl2r}.

\subsection{Phase shifts for fermionic fluctuations}\label{sec:fermionic}

The fermionic phase shifts can be directly extracted from the fermionic solution of a single giant magnon given in \cite{Minahan:2007gf,*Papathanasiou:2007gd}. However instead we will follow here the same route as in \cite{Chen:2007vs} and consider the dyonic giant magnon in terms of finite-gap solutions. Using the fact that the dyonic giant magnon only lives on the sphere part of $AdS_3\times S^3$ will be sufficient to determine the fermionic phase shifts  without knowing the precise form of the finite-gap equations. The consistency of the resulting finite-gap picture also serves as an additional check for the mixed-flux dyonic giant magnon results in \cite{Babichenko:2014yaa}.\\
\\
The equations of motion for strings on $AdS_3\times S^3$ correspond to the flatness condition of a current $j$ with an associated monodromy matrix $\Omega \sim P\exp\,(\oint j)$. Its analytic properties give rise to the finite-gap equations for classical solutions with periodic boundary conditions.\\
\\
In our case the monodromy matrix is an element of the supergroup $SU'(1,1|2)$ and its eigenvalues have the form
\begin{align}
  ( e^{ip_1^A(X)}, e^{ip_2^A(X)} | e^{ip_1^S(X)}, e^{ip_2^S(X)} ),\qquad p_1(X) = -p_2(X)
\end{align}
where the quasi-momenta $p_i(X)$ are complex functions of the spectral parameter $X$ and the label $A, S$ distinguishes between the $AdS_3$ and $S^3$ matrix blocks. These quasi-momenta have poles and branch cuts and thus can be written as
\begin{align}
 p_i(X) = G_i(X) + f_i(X)
\end{align}
where $f(X)$ contains the poles and the resolvent $G(X)$ contains the branch cuts with the discontinuity relation across each cut
\begin{align}
p_i(X+ i\epsilon) + p_j(X-i\epsilon) = 2\pi n_{ij},\quad n_{ij}\in \mathbb{Z}.
\end{align}
The resolvents characterise the different classical solutions and since we are only interested in solutions with non-trivial motion on the sphere the quasi-momenta take the form
\begin{align}
 p_1^A(X) = -p_2^A(X) = f(X),\qquad 
 p_i^S(X) = -p_2^S(X) = f(X) + G(X)  
\end{align}
In this formalism the phase shifts are encoded by introducing a micoscopic probe cut corresponding to small fluctuations. The associated discontinuity condition then becomes the quantisation condition of the associated plane wave momentum $k$ for a string solution of finite length $L$ 
\begin{align}
2\pi n_{ij} = p_i(X(k)+ i\epsilon) + p_j(X(k)-i\epsilon) = -\delta(k) -kL.
\end{align}
Each fluctuation corresponds to a connection of two particular sheets $(p_i^{\{S,A\}},p_j^{\{S,A\}})$ and their exact relation was determined in \cite{Gromov:2007aq}.\\
\\
In the plane wave limit the spectral parameter expands as $X\sim \ww \equiv \frac{r}{\qhat}$ and thus we have for $l=-1$
\begin{align}
 {p^A_-}_1(\ww_-) - {p^S_-}_2(\ww_-) &= 2G_-(\ww_-)  +2f_-(\ww_-) = -\delta_{Z_2}(\ww_-) -k(\ww_-)L\\
 {p^A_-}_1(\ww_-) - {p^A_-}_2(\ww_-) &=               2f_-(\ww_-) = -\delta_{Y_2}(\ww_-) -k(\ww_-)L
\end{align}
Since the phase shifts $\delta_{Y_2}$ vanish in the $AdS$ part we can identify the form of $f_-(x)$ as
\begin{align}
f_-(x) = -\frac{L}{2\qhat}\frac{2x}{(x+s)(x-s^{-1})}.
\end{align}
Furthermore we can read off the resolvent for the dyonic giant magnon from the bosonic phase shifts as
\begin{align}
G_-(x) = G_{mag}(x) - \frac{1}{2}G_{mag}(0), \quad G_{mag}(x) = i\ln \frac{x-X^+}{x-X^-},\quad G_{mag}(0) = p.
\end{align}
where $p$ is the magnon world-sheet momentum. For $l=1$ excitations we use \eqref{wwl2r} to get
\begin{align}
 {p^S_+}_1(\ww_+) - {p^S_+}_2(\ww_+) &= 2G_+(\ww_+)  +2f_+(\ww_+) = -\delta_{Z_2}\Big(\frac{1}{\ww_+}\Big) -k\Big(\frac{1}{\ww_+}\Big)L\\
 {p^A_+}_1(\ww_+) - {p^A_+}_2(\ww_+) &=               2f_+(\ww_+) = -\delta_{Y_2}\Big(\frac{1}{\ww_+}\Big) -k\Big(\frac{1}{\ww_+}\Big)L
\end{align}
and therefore
\begin{align}
f_+(x) = \frac{L}{2\qhat}\frac{2x}{(x-s)(x+s^{-1})},\qquad G_+(x) = G_{mag}\Big(\frac{1}{x}\Big) - \frac{1}{2}G_{mag}(0).
\end{align}
We see that the resolvent has the same form as in the $q=0$ case. Also these results for $f_\pm(x)$ and $G_\pm(x)$ match (C.1) in \cite{Babichenko:2014yaa} up to factors of $G_{mag}(0)$ which depend on the choice of boundary conditions for the dyonic giant magnon. In our case we have $Z_1\sim e^{it\pm ip/2}$ as $x\rightarrow\pm\infty$ corresponding to $- \frac{1}{2}G_{mag}(0)$ in both $G_+(x)$ and $G_-(x)$.\\
\\
Finally for the fermionic fluctuations we have for plane waves of $l=-1$ type
\begin{align}
{p_-^S}_1(\ww_-) - {p_-^A}_2(\ww_-) &=  G_-(\ww_-) +      2f_-(\ww_-) = -\delta_{\eta}(\ww_-)       -k(\ww_-)L\\
{p_-^S}_2(\ww_-) - {p_-^A}_1(\ww_-) &=  G_-(\ww_-) +      2f_-(\ww_-) = -\delta_{\theta}(\ww_-)     -k(\ww_-)L.
%
%
%
\end{align}
The equations for plane waves of type $l=1$ are then obtained using \eqref{wwl2r} and using the same arguments for conjugate fields we find altogether 
\begin{align}
 \delta_\eta(k) &= \delta_\theta(k) =  \frac{1}{2}\delta_{Z_2}(k),\qquad \delta_{\bar\eta}(k) = \delta_{\bar\theta}(k) = \frac{1}{2}\delta_{\bar Z_2}(k).
\end{align}

\subsection{Corrections to the dispersion relation and the dressing phase}\label{sec:1loopeval}


We can now evaluate the 1-loop corrections \eqref{deltaE} and \eqref{deltaTheta} using the phase shifts $\delta_I$ for a plane wave scattering off a single dyonic giant magnon explicitly given by
\begin{align}
AdS_3 &: \delta_{Y_k}=\delta_{\bar Y_k} =0\\
S^3   &: \delta_{Z_1}=\delta_{\bar Z_1} =0,\quad \delta_{Z_2,l}(k,X) = -2G(\ww_l(k)^{-l},X),\quad \delta_{\bar Z_2,l}(k,X) = 2G(\ww_{-l}(k)^l,X)\\
\mathrm{fermionic} &: \delta_{\theta,l}(k,X) = \delta_{\eta,l}(k,X) = -G(\ww_l(k)^{-l},X)\\
                   &: \delta_{\bar\theta,l}(k,X) = \delta_{\bar\eta,l}(k,X) = G(\ww_{-l}(k)^l,X)\\
G(\ww,X) &= i\Big(-\frac{1}{2}\ln\frac{X^+}{X^-} + \ln\frac{\ww-X^+}{\ww - X^-}\Big).
\end{align}
where the lower label $l$ specifies whether the phase shift is for a left or right moving plane wave of spectral parameter $w_l(k)$ given by \eqref{planewaveRel} (for example $\delta_{\bar Z_2,-}(k,X) = 2G(1/\ww_+(k),X)$). Here $X$ is the spectral parameter a dyonic giant magnon with $l_X=-1$ and we obtain the phase shift for $l_X=1$ using \eqref{Xl2r}.\\
\\
We immediately see that the 1-loop energy shift vanishes
\begin{align}
\Delta E = \frac{1}{2\pi}\int_{-\infty}^\infty\dif k\, \omega(k)\frac{\partial}{\partial k}\Big([\delta_{Z_2}-\delta_\theta-\delta_\eta]+[\delta_{\bar Z_2}-\delta_{\bar\theta}-\delta_{\bar\eta}]\Big) = 0.
\end{align}
For the 1-loop phase correction \eqref{deltaTheta}
\begin{align}
\Delta \Theta(p_1,p_2) &= \frac{1}{2\pi}\sum_{I=1}^{N_F}(-1)^{F_I}\int_{-\infty}^\infty \dif k\,\frac{\partial\delta_I(k;p_1)}{\partial k}\delta_I(k;p_2)\label{deltaTheta2}
\end{align}
we have the choice of evaluating the integral for a left or right moving plane wave. Both choices are equivalent since
%
%
\begin{align}
 \int_{-s^{-1}}^s \dif w_+\, \frac{\partial G(\frac{1}{w_+};X)}{\partial w_+}G\Big(\frac{1}{w_+};Y\Big)
 = \int_{-s}^{s^{-1}} \dif \ww_-\, \frac{\partial G(\ww_-;X)}{\partial \ww_-}G\Big(\ww_-;Y\Big).
\end{align}
This also implies that the integrals for conjugate and non-conjugate fields are in fact the same.
%
%
%
Let us also rewrite \eqref{deltaTheta} in a manifestly antisymmetric form as we expect from unitarity. Dropping a total derivative and noticing that under \eqref{Xl2r}
\begin{align}
G(w_l,X)\rightarrow -G(1/w_l,X)
\end{align}
we can write the 1-loop phase correction for same-type excitations $l_1=l_2 =l$ as
\begin{align}
\Delta \Theta_{ll}(p_1,p_2) = \frac{1}{\pi}\int_{-s^{-l}}^{s^l} \dif w\,\,\Big(
\frac{\partial G(w,X)}{\partial w}G(w,Y)
-(X\leftrightarrow Y)\Big),\\
\frac{\partial G(w,X)}{\partial w} = \frac{i}{w-X^+}-\frac{i}{w-X^-},\quad s=\frac{\sqrt{1+q}}{\sqrt{1-q}}.
\end{align}
We can further split this integral and perform the integration explicitly to give
\begin{align}
 \Delta \Theta_{ll}(p_1,p_2) = \chi_l(X^+,Y^+)-\chi_l(X^+,Y^-)-\chi_l(X^-,Y^+)+\chi_l(X^-,Y^-)\label{1loopF1}
\end{align}
where
\begin{align}
 \chi_l(X,Y) &= -\frac{1}{\pi}\int_{-s^{-l}}^{s^l}  \frac{\dif u}{u-X}\Big(-\frac{1}{2}\ln Y+\ln(u-Y)\Big)- (X\leftrightarrow Y),\\
 &= -\frac{1}{\pi}[I_l(X,Y)-I_l(Y,X)]\\
 I_l(X,Y) &= \ln\Big(\frac{X-s^l}{X+s^{-l}}\Big)\Big[\ln(X-Y)-\frac{1}{2}\ln Y\Big]-\mathrm{Li}_2\Big(\frac{X-s^l}{X-Y}\Big)+\mathrm{Li}_2\Big(\frac{X+s^{-l}}{X-Y}\Big)\label{Iint}.
\end{align}
For the case of mixed excitations, i.e. $l_1 = -l_2 = l$, we can use \eqref{Xl2r} directly on the integral \eqref{Iint}. However in order to make manifest its antisymmetry under $X\leftrightarrow Y$, $l_1\leftrightarrow l_2$ let us write the integral \eqref{deltaTheta2} as
\begin{align}
\Delta\bar \Theta_{l_1l_2}(p_1,p_2) = -\frac{1}{\pi}\int_{-s^{-l_1}}^{s^{l_1}} \dif w\,\,\Big(
\frac{\partial G(w,X)}{\partial w}G\Big(\frac{1}{w},Y\Big)
-(X\leftrightarrow Y,\, l_1\leftrightarrow l_2)\Big)
\end{align}
giving
\begin{align}
 \Delta \bar \Theta_{l_1l_2}(p_1,p_2) = \bar\chi_l(X^+,Y^+)-\bar\chi_l(X^+,Y^-)-\bar\chi_l(X^-,Y^+)+\bar\chi_l(X^-,Y^-)\label{1loopF2}\\
\bar\chi_{l}(X,Y)= -\frac{1}{\pi}(\bar I_l(X,Y)-\bar I_l(Y,X))\\
\bar I_l(X,Y) = -\frac{1}{2}\ln Y \ln\frac{X-s^{l}}{X+s^{-l}}+\ln(X-s^{l})\ln(Y-s^{-l})-\ln(X+s^{-l})\ln(Y+s^{l})\\
+\ln\Big(s^{-2l}\frac{X-s^{l}}{X+s^{-l}}\Big)\ln(1-XY) -\Li_2\Big(s^{-l}\frac{Y+s^{l}}{1-XY}\Big)+\Li_2\Big(-s^{l}\frac{Y-s^{-l}}{1-XY}\Big).
\end{align}
These 1-loop corrections to the bound-state S-matrix come directly from the elementary dressing phases. This is because the bound-state S-matrix does not receive contributions from the BDS factor (or its mixed excitation scattering equivalent) as was pointed out in \cite{Chen:2007vs}. In order to find the elementary dressing phases we notice that we can use the same arguments as in the tree-level case in section \ref{sec:afsphases} since the bound-state 1-loop corrections are of the form \eqref{elemForm}. Thus the elementary 1-loop dressing phases are given by the above expressions but with the dyonic giant magnon spectral parameters $X$, $Y$ replaced by the elementary magnon spectral parameters $x$, $y$.\\
\\
As a consistency check let us take the semiclassical limit $h\rightarrow\infty$ with $hp\,\,\mathrm{fixed}$ such that the shortening conditions \eqref{shortcond} are solved by
\begin{align}
x_l^\pm = x_l \pm \frac{i}{2}\alpha_l(x_l) + \mathcal{O}\Big(\frac{1}{\hh}\Big),\quad \alpha_l(x) = \frac{2x^2}{h\qhat (x-s^l)(x+s^{-l})},\\
m_l=1 + qlk,\quad x_l\equiv\frac{m_l+\sqrt{m_l^2+\qhat^2k^2}}{\qhat k}.
\end{align}
We obtain
\begin{align}
\Delta\Theta_{ll} = 2\theta_{1ll}(x,y) = -\frac{\alpha_l(x)\alpha_l(y)}{2\pi}\Big[\frac{1}{\qhat}\frac{(x+y)(1-\frac{1}{xy})-\frac{4ql}{\qhat}}{(x-s^l)(x+s^{-l})(y-s^l)(y+s^{-l})}\frac{x+y}{x-y}\notag\\
+\frac{2}{(x-y)^2}\ln\Big(\frac{y-s^l}{x-s^l}\frac{x+s^{-l}}{y+s^{-l}}\Big)\Big]\\
\Delta\bar \Theta_{l_1l_2}(p_1,p_2) = 2\bar \theta_{l_1l_2}(x,y) = -\frac{\alpha_{l_1}(x)\alpha_{l_2}(y)}{2\pi}\Big[\frac{1}{\qhat}\frac{(x-y)(1+\frac{1}{xy})-\frac{4ql_1}{\qhat}}{(x-s^{l_1})(x+s^{-{l_1}})(y+s^{-l_2})(y-s^{l_2})}\frac{1+xy}{1-xy}\notag\\
+\frac{2}{(1-xy)^2}\ln\Big(\frac{y-s^{l_2}}{x-s^{l_1}}\frac{x+s^{-l_1}}{y+s^{-l_2}}s^{l_1-l_2}\Big)\Big].
\end{align}
which matches the 1-loop dressing phase predictions in \cite{Babichenko:2014yaa,Engelund:2013fja,*Bianchi:2014rfa}.

\section{Concluding remarks}\label{sec:conclusions}

We have presented a semiclassical derivation of the tree-level and 1-loop dressing phases in the massive sector of string theory on $AdS_3 \times S^3 \times T^4$ with R-R and NS-NS 3-form fluxes. In both cases we have found agreement with the proposals from finite-gap equations and algebraic curve quantisation in \cite{Babichenko:2014yaa} and unitarity cut methods in \cite{Engelund:2013fja,*Bianchi:2014rfa}.\\
\\
For the 1-loop phase we have seen that the the semiclassical scattering data for bosonic fluctuations, given in terms of phase shifts experienced by plane waves scattering off dyonic giant magnons, is sufficient to determine the fermionic scattering data and the resolvent for the mixed-flux dyonic giant magnon which agrees with the suggested resolvent in \cite{Babichenko:2014yaa}.\\
\\
It would be interesting to extend these semiclassical methods to the massless sector if possible. The dispersion relation for massless modes has been found recently using off-shell symmetry algebra considerations in \cite{Lloyd:2014bsa}. Although a 2-loop near-flat space check, performed in \cite{Sundin:2014ema}, supports the exact form of the massive dispersion relation \eqref{dispRel} it disagrees in the massless case. %
Therefore an important question is whether we can provide a further check by identifying the analogue of the giant magnon for massless modes. There are two approaches we can take:\\

(i) One should be able to choose an ansatz living on the appropriate subspace of $AdS_3\times S^3 \times T^4$ which gives the dispersion relation we expect. However this might be the giant magnon itself which does not move on $T^4$ where the massless modes arise. This leads to the question why the giant magnon on $\R\times S^2$ cannot be interpreted as the giant magnon in the massless sector.\\

(ii) One might extend the mixed-flux dressing method used in this paper to include motion on $T^4$. In order to interpret the resulting solution as the giant magnon for massless modes we would expect non-trivial motion on the torus. This would allow us to construct scattering solutions and address the question of the semiclassical and 1-loop dressing phases in the massless and mixed mass sectors.

\section*{Acknowledgments}

AS would like to thank O. Ohlsson-Sax and A.A. Tseytlin for useful discussions. AS acknowledges the support of the STFC grant ST/J000353/1.

\appendix

\section{Time-delay from the complex sine-Gordon model}\label{app:csg}
 
The CsG soliton anti-soliton scattering solution for $q=0$ can be obtained from the single soliton solution by applying Bäcklund transformations and is given by (see e.g. \cite{Bowcock:2002vz})
\begin{align}
\psi_{2s} &= \frac{e^{i\gamma}(\delta_2\cc{v}_1-\delta_1\cc{v}_2)(\delta_1u_1-\delta_2u_2)- e^{-i\gamma}(\delta_2v_2-\delta_1v_1)(\delta_1u_2-\delta_2u_1) }{\delta_1^2-(u_2 \cc{u}_1+\cc{u}_2u_1+v_2\cc{v}_1+\cc{v}_2v_2)\delta_1\delta_2+\delta_2^2}
\end{align}
where
\begin{align}
 u_k &= \frac{N_k\cos\rho_k \exp(i\sin(\rho_k) {\cal T}_k)}{\cosh(\cos(\rho_k) {\cal X}_k)},\qquad 
 v_k = -e^{i\gamma} \Big(\cos(\rho_k)\tanh(\cos(\rho_k){\cal X}_k)+i\sin\rho_k\Big)\\
 \delta_k &= \exp(-\theta_k),\quad N_k = e^{i\phi_k} = \mathrm{const}\\
 {\cal X}_k &= \cosh(\theta_k)x - \sinh(\theta_k)t ,\qquad {\cal T}_k = \cosh(\theta_k)t- \sinh(\theta_k)x.
\end{align}
Note that $\gamma$ drops out of the full scattering solution $\psi_{2s}$. Considering the form of a single soliton solution for $q\ne 0$ in \cite{Hoare:2013lja} we can generalise $\psi_{2s}$ to $q\ne 0$ by performing the replacement
\begin{align}
 {\cal X}_k & \rightarrow \tilde{\cal X}_k = {\cal X}_k + q {\cal T}_k,\\
 {\cal T}_k & \rightarrow \tilde{\cal T}_k = {\cal T}_k + q {\cal X}_k.
\end{align}
One can easily check that this new solution satisfies the $q\ne 0$ CsG equation of motion
\begin{align}
\ddot{\psi} - \psi'' + \cc{\psi} \frac{\dot{\psi}^2-\acute{\psi}^2}{1-\lvert \psi \rvert^2} + (1-q^2)\psi (1-\lvert\psi\rvert^2) = 0.
\end{align}
Even though the $q\ne 0$ CsG model differs from the $q=0$ model only by a mass rescaling, the $q\ne 0$ soliton solution, which corresponds to the giant magnon, is not obtained through the simple mass rescaling $({\cal X},{\cal T}) \rightarrow \sqrt{1-q^2}({\cal X},{\cal T})$ but rather through a mixing of ${\cal X}$ and ${\cal T}$.\footnote{A simple mass rescaling in the CsG soliton solution would lead to embedding coordinates with $x,t$ rescaled breaking the Virasoro constraints.}\\
\\
The energy of this soliton is
\begin{align}
E = \int \dif x\, \mathcal{H} = \int_{-\infty}^{+\infty} \dif x\, \Big(\frac{\lvert\dot{\psi}\rvert^2+\lvert\acute{\psi}\rvert^2}{1-\lvert \psi\rvert^2}+(1-q^2)\lvert\psi\rvert^2\Big) = \sum_i 4(\cosh\theta_i - q\sinh\theta_i)\cos\rho_i
\end{align}
and we shall consider the COM frame where
\begin{align}
\sum_i E_i v_{si} = \sum_i (\sinh \theta_i- q \cosh \theta_i)\cos \rho_i = 0.
\end{align}
Taking the limit $t\rightarrow \infty$ with $x={v_s}_1t$ the free solitons shifted by $t\rightarrow t-\delta t_+$ become
\begin{align}
u_1 = e^{i(v_1 + \delta t_+\sin\rho_1(\cosh\theta_1-q\sinh\theta_1))}\cos\rho_1\sech\big[\delta t_+ \cos\rho_1(\sinh\theta_1-q\cosh\theta_1)\big],\,\,
u_2 = 0.
\end{align}
whereas the scattering soliton solution asymptotes to
\begin{align}
 \lvert \psi_{2s}\rvert^2 = \cos^2 \rho_1 \frac{1-\cos(2\rho_1)-\cos(2\rho_2)+\cosh(2(\theta_1-\theta_2))-4\cosh(\theta_1-\theta_2)\sin\rho_1\sin\rho_2}{2(\cosh(\theta_1-\theta_2)-\sin\rho_1\sin\rho_2)^2}.
\end{align}
Comparing these expression we obtain the expected result for the time-delay
\begin{align}
\Delta T = \frac{1}{(\sinh\theta_1-q\cosh\theta_1)\cos\rho_1}\Big\lvert\ln \frac{\cosh(\theta_1-\theta_2)-\cos(\rho_1-\rho_2)}{\cosh(\theta_1-\theta_2)+\cos(\rho_1+\rho_2)}\Big\rvert.
\end{align}

\parskip=0.pt
\baselineskip 11pt

\bibliographystyle{utphysM}
\bibliography{phase_v2_pubs}

\end{document}